\documentclass{jfm}
\usepackage{color}
\usepackage{amstext}
\usepackage{amssymb}
\usepackage{graphicx}
\usepackage{amsmath}

\newcommand\const{\mathrm{const}}
\newcommand\Div{\mathrm{div}\,}

\newcommand\vX{\boldsymbol{X}}

\newcommand\vV{\boldsymbol{V}}

\newcommand\vf{\boldsymbol{f}}
\newcommand\vh{\boldsymbol{h}}

\newcommand\vu{\boldsymbol{u}}
\newcommand\vv{\boldsymbol{v}}

\newcommand\vx{\boldsymbol{x}}

\newcommand\vg{\boldsymbol{g}}

\newcommand\vomega{\boldsymbol{\omega}}

\newcommand\vxi{\boldsymbol{\xi}}

\begin{document}

{\title[Theory of Non-Degenerated Oscillatory Flows] {Theory of Non-Degenerated Oscillatory Flows}}

\author[V. Vladimirov ]
{V.\ns A.\ns V\ls l\ls a\ls d\ls i\ls m\ls i\ls r\ls o\ls v}

\affiliation{Dept of Mathematics, University of York, Heslington, York, YO10 5DD, UK}

\pubyear{2011} \volume{xx} \pagerange{xx-xx}
\date{Oct 12th 2011 and in revised form ???}

\setcounter{page}{1}\maketitle \thispagestyle{empty}

\begin{abstract}

The aim of this paper is to derive the averaged governing equations for non-degenerated oscillatory flows, in
which the magnitudes of mean velocity and oscillating velocity are similar. We derive the averaged equations
for a scalar passive admixture, for a vectorial passive admixture (magnetic field in kinematic MHD), and for
vortex dynamics. The small parameter of our asymptotic theory is the inverse dimensionless frequency
$1/\sigma$. Our mathematical approach combines the two-timing method, distinguished limits, and the use of
commutators to simplify calculations. This approach produces recurrent equations for both the averaged and
oscillating parts of unknown fields. We do not use any physical or mathematical assumptions (except the most
common ones) and present calculations for the first three (zeroth, first, and second) successive
approximations. In all our examples the averaged equations exhibit the universal structure: the
Reynolds-stress-type terms (or the cross-correlations) are transformed into drift velocities,
pseudo-diffusion, and  two other terms reminiscent of Moffatt's mean-fields in turbulence. In particular, the
averaged motion of a passive scalar admixture is described only by a drift and pseudo-diffusion. The averaged
equations for a passive vectorial admixture and for vortex dynamics possess two mean-field terms, additional
to pseudo-diffusion. It is remarkable that all mean-field terms (including pseudo-diffusion) are expressed by
invariant operators (Lie-derivatives) which measure the deviation of some tensors from their `frozen-in'
values. Some physical assumptions and the results can be build upon obtained averaged equations. Our physical
interpretation suggests purely kinematic nature of pseudo-diffusion.

\end{abstract}

\section{Introduction \label{sect01}}

Oscillatory flows play important roles in all branches of fluid dynamics, especially in wave theory, in
biological and medical fluid mechanics, in geophysical and astrophysical fluid dynamics. If a chosen fluid
flow is not oscillating, it can be easily made so by adding some oscillations to boundary conditions,
\emph{etc}. There is no unique definition of an oscillatory flow. In this paper we suggest that such
a flow contains  velocity oscillations with  frequency higher than inverse characteristic times of all other
co-existing motions. It can be expressed as $\sigma\gg 1$, where $\sigma$ is dimensionless frequency of
oscillations. Our term \emph{a non-degenerated oscillatory flow} means that the mean  and oscillatory
velocity are similar in magnitudes. This class of flows is rather restrictive: in many applications the
oscillating part of velocity is dominating (hence, the flows are degenerated), see
\cite{Stokes,Craik0, Pedley, Riley, Buhler, VladimirovX1,VladimirovX2}.
There are four major incentives for the research described in this paper:
\newline
(1) To derive the averaged equations of  non-degenerated oscillatory flows. Such flows appear in various
applications and their study represents a necessary step towards the systematic studies of various classes of
degenerated flows.
\newline\noindent
(2) To make all the calculations and derivations elementary and free of any physical and mathematical
assumptions (except the most common ones, like the existence of a differentiable solutions). To use only
Eulerian description and  Eulerian averaging operation, since they are the most transparent and allows to
avoid additional steps and difficulties. For example, Eulerian description does not experience well-known
difficulties with chaotic trajectories. Such a deliberately simple framework would allow  to build further
physical models based on a solid ground.
\newline\noindent
(3) To study the  reduction of the Reynolds-stress-type terms to the drift-type terms, see
\cite{CraikLeib, Riley, VladimirovX1, VladimirovX2,Ilin}. This rather surprising reduction was discovered
by \cite{CraikLeib} in the small-amplitude theory of Langmuir circulations, but its generality and importance
in many other areas of fluid dynamics has not yet been recognized. In particular, it is interesting to find
what else (besides the drift velocities) can be produced by the Reynolds-stresses-type terms.
\newline\noindent
(4) To understand how universal is the structure of the averaged equations. In order to describe  universal
features and variations of transformations of the Reynolds-stresses-type terms (or  cross-correlations) we
study three different problems: (i)  a passive scalar admixture, (ii) a passive vectorial admixture (magnetic
field in kinematic MHD), and (iii) an active vectorial admixture (vortex dynamics). This target is
methodologically important: after establishing the universal building blocks of the theory one can  use them
in various degenerated cases; examples of such use can be found in
\cite{VladimirovX1,VladimirovX2}.

For the averaging and transforming of the governing equations we employ the two-timing method, see
\emph{e.g.} \cite{Nayfeh,Kevor}.  Our version of the method represents an elementary, systematic, and
justifiable procedure introduced by \cite{VladimirovPend, Yudovich, VladimirovDisc, VladimirovX1}. This
procedure is complemented by consideration of the distinguished limits, that develops  results of
\cite{VladimirovX1,VladimirovX2}. We also actively use the properties of commutators, which allow us to
simplify  analytical calculations. The employment of these methods and tools produces  recurrent equations
for both the averaged and oscillating parts of  unknown fields. In all problems (i)-(iii) we present the
calculations of three  successive approximations (zeroth, first, and second).

In \emph{Sect.}2 the notations are briefly introduced and the list the main definitions (of the averaging
operation,
\emph{etc.}) is presented.

In \emph{Sects.}3-5 we show that in all problems (i)-(iii) the averaged equations exhibit the universal
structure: the zeroth approximations repeat the structure of the original equations; in the first and second
approximation the Reynolds-stress-type terms are transformed into drift velocities and pseudo-diffusion,
which are the same for all three problems; two additional terms (reminiscent of the \emph{mean-field terms}
in turbulence, see \cite{Moffatt, Moffatt1}) also appear for magnetic fields and vortex dynamics. In
particular, the averaged motion of a passive scalar admixture is described only by a drift in the first
approximation, and by a combination of a drift and pseudo-diffusion in the second approximation.  It is
remarkable that all mean-field terms (including  pseudo-diffusion) are expressed by invariant operators
(Lie-derivatives) that measure the deviation of some tensors from their `frozen-in' values.

\emph{Sect.}6 is devoted to the notion of a distinguished limit. Here we explain why for the non-degenerated velocity
fields a slow time-variable $s$ coincides with physical time $t$. Such a coincidence allows us to build a
theory using only two time-variables $s=t$ and $\tau=\sigma t$.

\emph{Sect.}7  expands the area of applicability of all previously exposed results. We show that
for all three considered problems the high-frequency asymptotic solutions and the small-amplitude asymptotic
solutions can be mathematically identical. This property is valid for $s$-independent velocity fields.

\emph{Sect.}8 contains  physical analysis of the nature of pseudo-diffusion. Our consideration suggests
that this nature is purely kinematic,  it is not related to physical diffusion.

\emph{Sect.}9 (\emph{Discussion}) is devoted to the restrictions and limitations of the introduced transformations
of the governing equations, as well as to their possible generalizations.

\section{Used Functions and Notations}

The variables $\vx=(x_1,x_2,x_3)$, $t$, $s$, and $\tau$  in the text below serve as  dimensionless cartesian
coordinates, physical time, slow time, and fast time. The used definitions, notations, and properties are
itemized with bullets ($\bullet$):

\noindent
$\bullet$ The class $\mathbb{H}$ of \emph{hat-functions}  is defined as
\begin{eqnarray}
\widehat{f}\in \mathbb{H}:\quad
\widehat{f}(\vx, s, \tau)=\widehat{f}(\vx,s,\tau+2\pi)\label{tilde-func-def}
\end{eqnarray}
where the $\tau$-dependence is always $2\pi$-periodic; the dependencies on $\vx$ and $s$ are not specified.

\noindent
$\bullet$ The subscripts $t$, $\tau$, and $s$ denote the related partial derivatives.

\noindent
$\bullet$ For an arbitrary $\widehat{f}\in \mathbb{H}$ the \emph{averaging operation} is
\begin{eqnarray}
\langle {\widehat{f}}\,\rangle \equiv \frac{1}{2\pi}\int_{\tau_0}^{\tau_0+2\pi}
\widehat{f}(\vx, s, \tau)\,d\tau,\qquad\forall\ \tau_0\label{oper-1}
\end{eqnarray}
where  during the $\tau$-integration $s=\const$ and $\langle {\widehat{f}}\rangle$ does not depend on
$\tau_0$.

\noindent
$\bullet$  The class $\mathbb{T}$ of \emph{tilde-functions} is such that
\begin{eqnarray}
\widetilde f\in \mathbb{T}:\quad
\widetilde f(\vx, s, \tau)=\widetilde f(\vx,s,\tau+2\pi),\quad\text{with}\quad
\langle \widetilde f \rangle =0,\label{oper-2}
\end{eqnarray}
Tilde-functions are also called purely oscillating functions; they represent a special case of hat-functions
with the zero average.

\noindent
$\bullet$  The class $\mathbb{B}$ of \emph{bar-functions} is defined as
\begin{eqnarray}
\overline{f}\in \mathbb{B}:\quad  \overline{f}_{\tau}\equiv 0,\quad
\overline{f}(\vx, s)=\langle\overline f(\vx,s)\rangle
 \label{oper-3}
\end{eqnarray}

\noindent
$\bullet$  Any $\mathbb{H}$-function can be uniquely separated into its bar- and tilde- parts with the use of
(\ref{oper-1}):
\begin{eqnarray}
\widehat{f}=\overline{f}+\widetilde{f} \label{oper-3a}
\end{eqnarray}

\noindent
$\bullet$
\emph{The $\mathbb{T}$-integration (or the tilde-integration)}: for a given $\widetilde{f}$ we introduce a new function
$\widetilde{f}^{\tau}$ called the $\mathbb{T}$-integral of $\widetilde{f}$:
\begin{eqnarray}
&&\widetilde{f}^{\tau}\equiv\int_0^\tau \widetilde{f}(\vx,s,\rho)\,d\rho
-\frac{1}{2\pi}\int_0^{2\pi}\Bigl(\int_0^\mu
\widetilde{f}(\vx,s,\rho)\,d\rho\Bigr)\,d\mu\label{oper-7}
\end{eqnarray}

\noindent
$\bullet$ The unique solution of a PDE inside the tilde-class
\begin{eqnarray}
\widetilde{f}_\tau\equiv\partial \widetilde{f}/\partial\tau=0\quad \Rightarrow\quad \widetilde{f}\equiv 0\label{f-tilde=0}
\end{eqnarray}
follows from (\ref{oper-7}).

\noindent
$\bullet$ The $\tau$-derivative of a tilde-function always represents a tilde-function. However the
$\tau$-integration of a tilde-function can produce a hat-function. An example: let us take
$\widetilde\phi=\overline\phi_0\sin\tau$ where $\overline\phi_0$ be an arbitrary bar-function: one can see
that $\langle\widetilde\phi\,\rangle\equiv 0$, however
$\langle\int_0^\tau\widetilde\phi(\vx,s,\rho)d\rho\rangle=\overline\phi_0\neq 0$, unless
$\overline\phi_0\equiv 0$. Formula (\ref{oper-7}) keeps the result of integration inside the
$\mathbb{T}$-class.

$\bullet$ The $\mathbb{T}$-integration is inverse to the $\tau$-differentiation
$(\widetilde{f}^{\tau})_{\tau}=(\widetilde{f}_{\tau})^{\tau}=\widetilde{f}$; the proof is omitted.

\noindent
$\bullet$ The product of two tilde-functions $\widetilde{f}$ and $\widetilde{g}$ represents a hat-function:
$\widetilde{f}\widetilde{g}\equiv\widehat{F}$, say. Separating its tilde-part we write
\begin{eqnarray}
&&\widetilde{F}=\widehat{F}-\langle\widehat{F}\rangle
=\widetilde{f}\widetilde{g}-\langle\widetilde{f}\widetilde{g}\rangle=\{\widetilde{f}\widetilde{g}\}
\label{oper-5}
\end{eqnarray}
where the notation $\{\cdot\}$ is introduced to avoid two levels of tildes.

\noindent
$\bullet$ A dimensionless function $f=f(\vx,s,\tau)$ belongs to the class $\mathbb{O}(1)$
\begin{eqnarray}
f\in \mathbb{O}(1)\label{all-one}
\end{eqnarray}
if $f={O}(1)$ and all  partial $\vx$-, $s$-, and $\tau$-derivatives of $f$ (required for our consideration),
are also ${O}(1)$.

\noindent
$\bullet$ All large or small parameters in this paper are represented by various degrees of a large parameter
$\sigma$ (which represents dimensionless frequency) only; large or small parameters appear as explicit
multipliers in all formulae, while all functions always belong to $\mathbb{O}(1)$-class.

\noindent
$\bullet$ The commutator of two vector-fields $\vf$ and $\vg$ is
\begin{eqnarray}
&& [\vf,\vg]\equiv(\vg\cdot\nabla)\vf-(\vf\cdot\nabla)\vg,\label{commutr}
\end{eqnarray}
It is antisymmetric and satisfies Jacobi's identity for any vector-fields $\vf$, $\vg$, and $\vh$:
\begin{eqnarray}
&&[\vf,\vg]=-[\vg,\vf],\quad [\vf,[\vg,\vh]]+ [\vh,[\vf,\vg]]+[\vg,[\vh,\vf]]=0 \label{oper-13}
\end{eqnarray}
$\bullet$ From (\ref{oper-3a}) and (\ref{oper-3}) one can see that
\begin{eqnarray}
\widehat{f}_\tau=\widetilde{f}_\tau, \quad
\langle\widehat{f}_\tau\rangle=\langle\widetilde{f}_\tau\rangle=0 \label{oper-6}
\end{eqnarray}
$\bullet$ As the average operation (\ref{oper-1}) is proportional to the integration over $\tau$, the
integration by parts yields
\begin{eqnarray}
&&\langle\widetilde{f}\widetilde{g}_\tau\rangle=\langle(\widetilde{f}\widetilde{g})_\tau\rangle-
\langle\widetilde{f}_\tau\widetilde{g}\rangle=-\langle\widetilde{f}_\tau \widetilde{g}\rangle=-
\langle\widetilde{f}_\tau \widehat{g}\rangle
\label{oper-9}\\
&&\langle\widetilde{f}_\tau\widetilde{g}\widetilde{h}\rangle+\langle\widetilde{f}\widetilde{g}_\tau\widetilde{h}\rangle+
\langle\widetilde{f}\widetilde{g}\widetilde{h}_\tau\rangle=0
\label{oper-9a}\\
&&\langle\widetilde{f}\widetilde{g}^\tau\rangle=\langle(\widetilde{f}^\tau\widetilde{g}^\tau)_\tau\rangle-
\langle\widetilde{f}^\tau\widetilde{g}\rangle=-\langle\widetilde{f}^\tau \widetilde{g}\rangle=-
\langle\widetilde{f}^\tau \widehat{g}\rangle
\label{oper-10}\\
&&\langle[\widetilde{\vf},\widetilde{\vg}_\tau]\rangle=-\langle[\widetilde{\vf}_\tau,\widetilde{\vg}]\rangle=-
\langle[\widetilde{\vf}_\tau, \widehat{\vg}]\rangle,\
\langle[\widetilde{\vf},\widetilde{\vg}^\tau]\rangle=-\langle[\widetilde{\vf}^\tau,\widetilde{\vg}]\rangle=-
\langle[\widetilde{\vf}^\tau, \widehat{\vg}]\rangle
\label{oper-15}
\end{eqnarray}

Many of the above definitions and terms are different from the ones used in various branches of physics and
fluid dynamics. We have introduced our own terms in order to avoid ambiguities.

\section{Scalar Admixture}

\subsection{Two-timing formulation}

The equation for a passive scalar admixture $c$ transported by a given velocity field  $\vu^*$ in an
incompressible fluid is
\begin{eqnarray}\label{scalar-0-1}
&&\frac{\partial{c}}{\partial {t}^*}+\vu^*\cdot\nabla^*{c}=0,\quad \nabla^*\cdot\vu^*=0
\end{eqnarray}
where asterisks mark dimensional variables, ${t}^*$-time, $\vx^*=(x_1^*,x_2^*,x_3^*)$-cartesian coordinates,
$\nabla^*=(\partial/\partial x_1^*, \partial/\partial x_2^*,\partial/\partial x_3^*)$. This paper id focused
on the transformations of the governing equations, therefore the geometry a flow domain and particular
boundary conditions can be introduced at later stages of the research (see Discussion). We accept that the
considered class of oscillatory flows $\vu^*=\vu^*(\vx^*,t^*;\sigma^*)$ possesses independent characteristic
scales of velocity $U$, length $L$, time $T$, and frequency $\sigma^*$
\begin{eqnarray}
&& U,\quad L,\quad T,\quad \sigma^*;\quad T_i\equiv L/U,\quad T=T_i
\label{scalar-0-2}
\end{eqnarray}
where $T_i$ is a dependent (intrinsic) time-scale. It will be shown by the distinguished limit consideration
that for non-degenerated flows the scales $T$ and $T_i$ are of similar order (one can accept that $T=k T_i$
with a constant $k\sim 1$; however without any restriction of generality this constant can be taken as $k=
1$). Then dimensionless (not asteriated) variables and frequency are
\begin{eqnarray}
&& \vx\equiv\vx^*/L,\quad t^*\equiv t/T,
\quad\vu\equiv\vu^*/U,\quad\sigma\equiv\sigma^*T;\quad 1/\sigma\ll 1
\label{scalar-0-3}
\end{eqnarray}
where $1/\sigma$ is our basic small parameter. Due to the linearity of (\ref{scalar-0-1}) we consider the
scalar field $c$  being dimensionless. The dimensionless version of (\ref{scalar-0-1}) is
\begin{eqnarray}\label{scalar-0-4}
&&{c}_{t}+\vu\cdot\nabla{c}=0
\end{eqnarray}
The \emph{given} oscillating velocity $\vu$ is taken as a hat-function (\ref{tilde-func-def})
\begin{eqnarray}
\vu=\widehat{\vu}(\vx,t, \sigma t)\quad\text{or}\quad \vu=\widehat{\vu}(\vx,s, \tau)\quad  \text{with}\, \ \tau=\sigma {t},\ s= t
\label{scalar-0-5}
\end{eqnarray}
where $s$ and $\tau$ are two \emph{mutually dependent} time-variables ($s$ is \emph{slow time} and $\tau$ is
\emph{fast time}). We assume that the solution of (\ref{scalar-0-4}),(\ref{scalar-0-5}) also represents a
hat-function
\begin{eqnarray}
&& {c}=\widehat{{c}}(\vx, s, \tau)\label{scalar-0-6}
\end{eqnarray}
In order to justify this assumption one needs to build such an unique solution for  arbitrary initial data.
The chain rule transforms (\ref{scalar-0-4}) to
\begin{eqnarray}
&&\left(\frac{\partial}{\partial\tau}+
\frac{1}{\sigma}\frac{\partial}{\partial s}\right)\widehat{{c}}+
\frac{1}{\sigma}\widehat{\vu}\cdot\nabla\widehat{{c}}= 0\label{scalar-0-7a}
\end{eqnarray}
Eqn.(\ref{scalar-0-7a}) contains the only small parameter
\begin{eqnarray}
\varepsilon\equiv \frac{1}{T\sigma^*}=\frac{1}{\sigma}\label{scalar-0-8}
\end{eqnarray}
Here one must make an auxiliary (but technically essential) step: after the use of the chain rule
(\ref{scalar-0-7a}) the variables $s$ and $\tau$  are  considered to be (temporarily) \emph{mutually
independent}:
\begin{eqnarray}
&& \tau\quad\text{and}\quad s \quad \text{are independent variables (temporarily)}\label{scalar-0-9}
\end{eqnarray}
From the mathematical viewpoint the rise of the number of independent variables in a PDE represents a very
radical step, which leads to an entirely new PDE. This step has to be justified \emph{a posteriori} by
showing that the error of the obtained solution (rewritten back to the original variable $t$ and substituted
into the original equation (\ref{scalar-0-4})), is small.

Hence there is the equation which is going to be considered in detail
\begin{eqnarray}
&&\widehat{c}_\tau+
\varepsilon (\widehat{\vu}\cdot\nabla)\widehat{c}+\varepsilon\widehat{c}_s= 0,\quad \Div \widehat{\vu}=0
\quad \varepsilon\equiv 1/\sigma\to 0\label{scalar-main-eq}
\end{eqnarray}

\subsection{Results for the successive approximations of scalar field}

We look for the solutions of (\ref{scalar-main-eq}) in the form of regular series
\begin{eqnarray}
&&(\widehat{c},\widehat{\vu})=\sum_{k=0}^\infty\varepsilon^k (\widehat{c}_k,\widehat{\vu}_k);\quad
\widehat{c}_k, \widehat{\vu}_k\in \mathbb{H}\cap\mathbb{O}(1),\quad k=0,1,2,\dots
\label{scalar-2}
\end{eqnarray}
where all $\widehat{\vu}_k$ are given. The main approximation of the general velocity field $\widehat{\vu}$
is not degenerated:
\begin{eqnarray}
&&\widehat{\vu}_0=\overline{\vu}_0+\widetilde{\vu}_0,
\quad\text{with}\quad\overline{\vu}_0\not\equiv
0\quad\text{and}\quad\widetilde{\vu}_0\not\equiv 0
\label{scalar-0}
\end{eqnarray}
The substitution of (\ref{scalar-2}) into (\ref{scalar-main-eq}) produces the equations for the first three
successive approximations
\begin{eqnarray}
&&\widehat{c}_{0\tau}=0 \label{scalar-r-1}\\
&&\widehat{{c}}_{1\tau}+\widehat{\vu}_0\cdot\nabla \widehat{{c}}_0+\widehat{{c}}_{0s}=0\label{scalar-r-2}\\
&&\widehat{{c}}_{2\tau}+\widehat{\vu}_1\cdot\nabla \widehat{{c}}_0+
\widehat{\vu}_0\cdot\nabla\widehat{{c}}_1+\widehat{{c}}_{1s}=0
\label{scalar-r-3}\\
&&\widehat{{c}}_{3\tau}+\widehat{\vu}_2\cdot\nabla \widehat{{c}}_0+
\widehat{\vu}_0\cdot\nabla\widehat{{c}}_2+\widehat{\vu}_1\cdot\nabla\widehat{{c}}_1+
\widehat{{c}}_{2s}=0
\label{scalar-r-3a}\\
&&\widehat{c}_{k}=\overline{c}_k(\vx,t)+\widetilde{c}_k(\vx,t,\tau),\ \overline{c}_k\in \mathbb{B}\cap
\mathbb{O}(1),\
\widetilde{c}_k\in \mathbb{T}\cap \mathbb{O}(1), \ k=0,1,2,3\nonumber
\end{eqnarray}
The detailed solution/transformation of (\ref{scalar-r-1})-(\ref{scalar-r-3a}) is given in \emph{Appendix A}.
Here we formulate the result. The truncated general solution $\widehat{c}^{[3]}$ is
\begin{eqnarray}
&&\widehat{c}^{[3]}=\widehat{c}_0+\varepsilon\widehat{c}_1+\varepsilon^2\widehat{c}_2+\varepsilon^3\widehat{c}_3
\label{gs-0}
\end{eqnarray}
Its bar-parts $\overline{c}_k$ satisfy the equations
\begin{eqnarray}
&&\overline{{c}}_{0s}+\overline{\vu}_0\cdot\nabla \overline{{c}}_0=0\label{gs-7}\\
&&\overline{{c}}_{1s}+\overline{\vu}_0\cdot\nabla\overline{{c}}_1+
(\overline{\vu}_1+\overline{\vV}_0)\cdot\nabla\overline{{c}}_0 =0\label{gs-8}\\
&&\overline{{c}}_{2s}+\overline{\vu}_0\cdot\nabla\overline{{c}}_2+
(\overline{\vu}_1+\overline{\vV}_0)\cdot\nabla\overline{{c}}_1+\label{gs-9}\\
&&(\overline{\vu}_2+\overline{\vV}_{10}+\overline{\vV}_1+\overline{\vV}_{\chi})\cdot\nabla\overline{{c}}_1
=\frac{\partial}{\partial x_i}\left(
\overline{\chi}_{ik}\frac{\partial\overline{c}_0}{\partial x_k}\right)\nonumber
\\
&&\overline{\vV}_0\equiv \frac{1}{2}\langle[\widetilde{\vu}_0,\widetilde{\vxi}]\rangle,\quad
\overline{\vV}_1\equiv\frac{1}{3}\langle[[\widetilde{\vu}_0,\widetilde{\vxi}],\widetilde{\vxi}]\rangle,\quad
\widetilde{\vxi}\equiv\widetilde{\vu}_0^\tau
\label{gs-10}\\
&&\overline{\vV}_{10}\equiv\langle[\widetilde{\vu}_1,\widetilde{\vxi}]\rangle,
\quad
\vV_\chi\equiv\frac{1}{2}\langle[\widetilde{\vxi},\mathfrak{L}\widetilde{\vxi}]\rangle,\quad
\overline{\chi}_{ik}\equiv\frac{1}{2}
\mathfrak{L}\langle\widetilde{\xi}_i\widetilde{\xi}_k\rangle,\label{gs-13}
\end{eqnarray}
where the operator $\mathfrak{L}$ (Lie-derivative) is such that its action on $\widetilde{\vxi}$ is defined
as
\begin{eqnarray}
&&\mathfrak{L}\widetilde{\vxi}\equiv\widetilde{\vxi}_s+[\widetilde{\vxi},\overline{\vu}_0]\label{gs-13a}
\end{eqnarray}
and its action on any (constructed from $\widetilde{\vxi}$) tensorial field $f_{ik\dots}$ is such that
$\mathfrak{L}\widetilde{\vxi}=0$ implies $\mathfrak{L}f_{ik\dots}=0$. In particular,
\begin{eqnarray}
&&\mathfrak{L}\langle\widetilde{\xi}_i\widetilde{\xi}_k\rangle\equiv\left(\partial_s+
\overline{\vu}_0\cdot\nabla\right)\langle\widetilde{\xi}_i\widetilde{\xi}_k\rangle
-\frac{\partial\overline{u}_{0k}}{\partial x_m}\langle\widetilde{\xi}_i\widetilde{\xi}_m\rangle-
\frac{\partial\overline{u}_{0i}}{\partial x_m}\langle\widetilde{\xi}_k\widetilde{\xi}_m\rangle
\label{gs-14}
\end{eqnarray}
Formulae (\ref{gs-13a}) and (\ref{gs-14}) are also known as `frozen-in' operators for a vectorial field and
tensorial field. After (\ref{gs-7})-(\ref{gs-9}) are solved, the tilde-parts $\widetilde{c}_k$ of
(\ref{gs-0}) are given by the recurrent expressions
\begin{eqnarray}
&&\widetilde{c}_0\equiv 0,\label{gs-1}\\
&& \widetilde{{c}}_1=-\widetilde{\vxi}\cdot\nabla\overline{{c}}_0,\label{gs-2}\\
&&\widetilde{{c}}_{2}=-\widetilde{\vu}_1^\tau\cdot\nabla\overline{{c}}_0-
\overline{\vu}_0\cdot\nabla\widetilde{{c}}_1^\tau-\widetilde{\vxi}\cdot\nabla\overline{{c}}_1
-\{\widetilde{\vu}_0\cdot\nabla\widetilde{{c}}_1\}^\tau-\widetilde{{c}}_{1s}^\tau, \label{gs-3}\\
&&\widetilde{{c}}_{3\tau}=-\widetilde{\vu}_2^{\tau}\cdot\nabla\overline{{c}}_0-
\widetilde{\vu}_0^{\tau}\cdot\nabla\overline{{c}}_2-
\overline{\vu}_0\cdot\nabla\widetilde{{c}}_2^{\tau}-
\widetilde{\vu}_1^{\tau}\cdot\nabla\overline{{c}}_1-
\overline{\vu}_1\cdot\nabla\widetilde{{c}}_1^{\tau}-\label{gs-4}\\
&&\{\widetilde{\vu}_0\cdot\nabla\widetilde{{c}}_2\}^{\tau}-
\{\widetilde{\vu}_1\cdot\nabla\widetilde{{c}}_1\}^{\tau}
-\widetilde{{c}}_{2s}^{\tau}\nonumber
\end{eqnarray}

Three averaged equations (\ref{gs-7})-(\ref{gs-9}) can be written as a single advection-pseudo-diffusion
equation (with the error ${O}(\varepsilon^3)$)
\begin{eqnarray}
&&\left(\partial_s+ \overline{\vv}\cdot\nabla\right)\overline{c} =
\frac{\partial}{\partial x_i}\left(\overline{\kappa}_{ik}\frac{\partial\overline{c}}{\partial x_k}\right)
\label{gs-15}\\
&&\overline{c}=\overline{c}^{[2]}=\overline{c}_0+\varepsilon\overline{c}_1+\varepsilon^2\overline{c}_2,\quad
\overline{\vv}=\overline{\vu}^{[2]}+\overline{\vV}^{[2]}\label{gs-16}\\
&&\overline{\vu}^{[2]}=\overline{\vu}_0+\varepsilon\overline{\vu}_1+\varepsilon^2
\overline{\vu}_2,\quad \overline{\vV}^{[2]}=\varepsilon\overline{\vV}_0+\varepsilon^2
(\overline{\vV}_{10}+\overline{\vV}_1+\overline{\vV}_{\chi})\label{4.22}\\
&&\overline{\kappa}_{ik}=\overline{\chi}_{ik}^{[2]}=\varepsilon^2\overline{\chi}_{ik}\label{4.23}
\end{eqnarray}
where all the coefficients are given in (\ref{gs-10}),(\ref{gs-13}). Eqn. (\ref{gs-15}) shows that (with a
given precision ${O}(\varepsilon^2$)) the motion of $\overline{c}$ represents an advection with  velocity
$\overline{\vv}$ and pseudo-diffusion with  matrix coefficient
$\overline{\kappa}_{ik}=\varepsilon^2\overline{\chi}_{ik}$.

One can write the truncated solution
\begin{eqnarray}
&&\widehat{c}^{[3]}=\overline{c}_0+\overline{c}_1+\widetilde{c}_1+\overline{c}_2+\widetilde{c}_2+
\overline{c}_3+\widetilde{c}_3
\label{trunc-c3}
\end{eqnarray}
where all the terms except $\overline{c}_3$ are determined by the above equations. After rewriting
(\ref{trunc-c3}) back to the original variable $t$ ($s=t,\tau=\sigma t$) and its substitution into
(\ref{scalar-0-4}) one can show that the error term in the RHS of the equation is $O(\varepsilon^3)$. This
estimation might be considered as  \emph{mathematical justification} of the used procedure, including its
most sensitive part (\ref{scalar-0-9}).

\section{Kinematic MHD-equations}

\subsection{Two-timing formulation of the kinematic MHD}

The equation for a magnetic field $\vh$ to be `frozen' into a given  oscillating incompressible velocity
field $\vu^*(\vx^*,t^*;\sigma^*)$ is
\begin{eqnarray}\label{exact-1-h}
&&\frac{\partial\vh}{\partial {t}^*}+[\vh,\vu^*]^*=0,\quad \nabla^*\cdot\vu^*=0,\quad
\nabla^*\cdot\vh=0
\end{eqnarray}
where all the notations are the same as in \emph{Sect.}3.1;  $[\,\cdot\, ,
\cdot\,]^*$ stands for  dimensional commutator (\ref{commutr}), and $\vh$ is taken dimensionless due to linearity.
The analysis of dimensions is the same as in \emph{Sect.}3.1; the dimensionless version of (\ref{exact-1-h})
is
\begin{eqnarray}\label{exact-1-h-dl}
&&\vh_t+[\vh,\vu]=0,\quad \nabla\cdot\vu=0,\quad
\nabla\cdot\vh=0
\end{eqnarray}
The \emph{given} oscillating velocity is taken  as a hat-function (\ref{tilde-func-def})
\begin{eqnarray}
\vu=\widehat{\vu}(\vx,s, \tau);\quad  \text{with}\, \ \tau\equiv\sigma {t},\, s\equiv {t}
\label{given-u}
\end{eqnarray}
and the solution of (\ref{exact-1-h-dl}),(\ref{given-u}) is also taken as a hat-function
\begin{eqnarray}
&& \vh=\widehat{\vh}(\vx, s, \tau)\label{exact-2}
\end{eqnarray}
Then the following equation has to be studied
\begin{eqnarray}
&&\widehat{\vh}_\tau+ \varepsilon[\widehat{\vh},\widehat{\vu}]+\varepsilon\widehat{\vh}_s= 0,
\quad \varepsilon\to 0\label{main-eq-0-h}
\end{eqnarray}
where the variables $\tau$ and $s$ are again temporarily independent (\ref{scalar-0-9}).

\subsection{Results for the successive approximations of a magnetic field}

We look for the solution of (\ref{main-eq-0-h}) in the form of regular series
\begin{eqnarray}
&&(\widehat{\vh},\widehat{\vu})=\sum_{k=0}^\infty\varepsilon^k (\widehat{\vh}_k,\widehat{\vu}_k);\quad
\widehat{\vh}_k, \widehat{\vu}_k\in \mathbb{H}\cap\mathbb{O}(1),\quad k=0,1,2,\dots
\label{basic-4aa-0-h}
\end{eqnarray}
where all $\widehat{\vu}_k$ are given, and (as in \emph{Sect.}3.2) we consider a prescribed non-degenerated
velocity field: $\overline{\vu}_0\not\equiv 0$ and $\widetilde{\vu}_0\not\equiv 0$. The substitution of
(\ref{basic-4aa-0-h}) into (\ref{main-eq-0-h}) produces the equations for the first three successive
approximations
\begin{eqnarray}
&&\widehat{\vh}_{0\tau}=0 \label{eqn-0-0-h-f}\\
&&\widehat{\vh}_{1\tau}+[\,\widehat{\vh}_0,\widehat{\vu}_0]+\widehat{\vh}_{0s}=0\label{eqn-1-0-f}\\
&&\widehat{\vh}_{2\tau}+\widehat{\vh}_{1s}+[\widehat{\vh}_0,\widehat{\vu}_1]+
[\widehat{\vh}_1,\widehat{\vu}_0]=0\label{eqn-2-0-h-f}\\
&&\widehat{\vh}_{3\tau}+\widehat{\vh}_{2s}+[\widehat{\vh}_0,\widehat{\vu}_2]+
[\widehat{\vh}_2,\widehat{\vu}_0]+[\widehat{\vh}_1,\widehat{\vu}_1]=0\label{eqn-3-h-f}\\
&&\widehat{\vh}_{k}=\overline{\vh}_k(\vx,t)+\widetilde{\vh}_k(\vx,t,\tau),\ \overline{\vh}_k\in
\mathbb{B}\cap
\mathbb{O}(1),\
\widetilde{\vh}_k\in \mathbb{T}\cap \mathbb{O}(1), \ k=0,1,2,3\nonumber
\end{eqnarray}
The detailed solution/transformation of (\ref{eqn-0-0-h-f})-(\ref{eqn-3-h-f}) is given in
\emph{Appendix B}. Here we formulate the result. The truncated general solution $\widehat{\vh}^{[3]}$ is
\begin{eqnarray}
&&\widehat{\vh}^{[3]}=\widehat{\vh}_0+\varepsilon\widehat{\vh}_1+\varepsilon^2\widehat{\vh}_2+\varepsilon^3\widehat{\vh}_3
\label{gs-0}
\end{eqnarray}
The bar-parts $\overline{\vh}_k$ satisfy the equations
\begin{eqnarray}
&&\overline{\vh}_{0s}+[\,\overline{\vh}_0,\overline{\vu}_0]=0\label{eqn-1-0-bar-h-f}\\
&&\overline{\vh}_{1s}+ [\overline{\vh}_0,\overline{\vu}_1+\overline{\vV}_0]
+[\overline{\vh}_1,\overline{\vu}_0]=0
\label{eqn-3-bar1-h-f}\\
&&\overline{\vh}_{2s}+[\overline{\vh}_2,\overline{\vu}_0]+[\overline{\vh}_1,(\overline{\vu}_1+\overline{\vV}_0)]
+ [\overline{\vh}_0,(\overline{\vu}_2+\overline{\vV}_{10}+\overline{\vV}_1+\overline{\vV}_{\chi})] =
\label{eqn-3-bar1-h-f}\\
&&\frac{\partial}{\partial x_i}\left(
\overline{\chi}_{ik}\frac{\partial\overline{h}_{0i}}{\partial x_k}\right)-
\overline{b}_{ikl}\frac{\partial\overline{h}_{0l}}{\partial x_k}-\overline{a}_{ik}\overline{h}_{k}
\nonumber
\end{eqnarray}
were the notations are the same as in (\ref{gs-10}),(\ref{gs-13}) and
\begin{eqnarray}
\overline{b}_{ikl}\equiv \mathfrak{L}\left\langle\widetilde{\xi}_k\frac{\partial \widetilde{\xi}_i}
{\partial x_l}\right\rangle,\quad
\overline{a}_{ik}\equiv \mathfrak{L} \left\langle \widetilde{\xi}_l\frac{\partial^2 \widetilde{\xi}_i}
{\partial x_l\partial x_k}  -
\frac{\partial \widetilde{\xi}_l}{\partial x_k}\frac{\partial \widetilde{\xi}_i}{\partial x_l}\right\rangle
\label{h-18-f}
\end{eqnarray}
The operator $\mathfrak{L}$ is Lie-derivative, see (\ref{gs-13a}),(\ref{gs-14}). After
(\ref{eqn-1-0-bar-h-f})-(\ref{eqn-3-bar1-h-f}) are solved, the tilde-parts $\widetilde{\vh}_k$ of
(\ref{gs-0}) are given by the recurrent expressions
\begin{eqnarray}
&&\widetilde{\vh}_0\equiv 0,\label{sol-0-0-h-f}\\
&& \widetilde{\vh}_1=[\widetilde{\vu}_0^\tau,\overline{\vh}_0]\label{sol-1-0-h-f}\\
&&\widetilde{\vh}_{2}=-\widetilde{\vh}_{1s}^\tau-[\overline{\vh}_0,\widetilde{\vu}_1^\tau]-
[\widetilde{\vh}_1^\tau,\overline{\vu}_0]-[\overline{\vh}_1,\widetilde{\vu}_0^\tau]
-\{[\widetilde{\vh}_1,\widetilde{\vu}_0]\}^\tau
\label{eqn-2-0-tilde-h-integr-f}
\\
&&\widetilde{\vh}_{3}=-\widetilde{\vh}_{2s}^\tau-[\overline{\vh}_0,\widetilde{\vu}_2^\tau]-
[\widetilde{\vh}_2^\tau,\overline{\vu}_0]-[\overline{\vh}_2,\widetilde{\vu}_0^\tau]-\label{eqn-3-tilde-h-tau-f}\\
&&[\widetilde{\vh}_1^\tau,\overline{\vu}_1]-[\overline{\vh}_1,\widetilde{\vu}_1^\tau]-
\{[\widetilde{\vh}_2,\widetilde{\vu}_0]\}^\tau-
\{[\widetilde{\vh}_1,\widetilde{\vu}_1]\}^\tau\nonumber
\end{eqnarray}
Three averaged equations (\ref{eqn-1-0-bar-h-f})-(\ref{eqn-3-bar1-h-f}) can be written as a single equation
that combines advection, pseudo-diffusion, and other mean-field-type terms (with the error
${O}(\varepsilon^3)$)
\begin{eqnarray}
&&\overline{\vh}_s+[\overline{\vh},\overline{\vv}] =
\frac{\partial}{\partial x_k}\left(
\overline{\kappa}_{kl}\frac{\partial\overline{h}_{0i}}{\partial x_l}\right)-
\overline{B}_{ikl}\frac{\partial\overline{h}_{0l}}{\partial x_k}-\overline{A}_{ik}\overline{h}_{k}
\label{gh-15}\\
&&\overline{\vh}=\overline{\vh}^{[2]}=\overline{\vh}_0+\varepsilon\overline{\vh}_1+\varepsilon^2\overline{\vh}_2
\label{gh-16}
\end{eqnarray}
The notation for $\overline{\vv}$, $\overline{\vu}^{[2]}$, $\overline{\vV}^{[2]}$, and
$\overline{\kappa}_{ik}$ are the same as in (\ref{gs-16}), while $\overline{B}_{ikl}\equiv
\varepsilon^2\overline{b}_{ikl}$ and $\overline{A}_{ik}\equiv\varepsilon^2\overline{a}_{ik}$. Eqn. (\ref{gh-15})
shows that the evolution of $\overline{\vh}$ represents (with the error ${O}(\varepsilon^3)$) an advection
and stretching by  velocity $\overline{\vv}$, pseudo-diffusion with  matrix coefficient
$\overline{\kappa}_{ik}=\varepsilon^2\overline{\chi}_{ik}$, and other mean-field deformations. The terms with
$\overline{B}_{ikl}$ and $\overline{A}_{ik}$ are essentially new in comparison with (\ref{gs-15}). The
construction and justification of the truncated solution can be performed in a similar to
\emph{Sect.}3 way.

%
%

\section{Euler's equations}

\subsection{Two-timing problem of vorticity dynamics}

The governing equation for  dynamics of an inviscid incompressible fluid with velocity field $\vu^*$ and
vorticity $\vomega^*$ is taken in the vorticity form
\begin{eqnarray}
&&\frac{\partial\vomega^*}{\partial {t}^*}+[\vomega^*,\vu^*]^*=0,\quad \quad \nabla^*\cdot\vu^*=0\label{exact-1v}\\
&&\vomega^*\equiv\nabla^*\times\vu^*\label{exact-1vv}
\end{eqnarray}
where all the notations are the same as in \emph{Sects.}3.1 and 4.1. The equations (\ref{exact-1v}) are
mathematically the same as (\ref{exact-1-h}) in MHD-kinematics. However, they are complemented by an
additional constraint (\ref{exact-1vv}). As a result the velocity field represents an unknown function and
can not be considered as the prescribed one. It is remarkable, that this very sound difference does not have
any impact on all the results of\emph{ Sect.}4 and the derivations of \emph{Appendix B}. The sufficient (for
the derivation of the averaged equations for vorticity) operations are: we have to replace $\vh$ by $\vomega$
and write the constraint (\ref{exact-1vv}) additionally to all equations. Of course, the same averaged
equations can be obtained independently.

The  system of dimensionless equations under consideration can be written as:
\begin{eqnarray}
&&\widehat{\vomega}_\tau+ \varepsilon[\widehat{\vomega},\widehat{\vu}]+\varepsilon\widehat{\vomega}_s= 0,
\quad \widehat{\vomega}=\nabla\times\widehat{\vu},\quad \Div\widehat{\vu}=0;\quad \varepsilon\to 0\label{main-eq-0}
\end{eqnarray}
We are looking for the solution of these equations in the form of regular series
\begin{eqnarray}
&&(\widehat{\vomega},\widehat{\vu})=\sum_{k=0}^\infty\varepsilon^k(\widehat{\vomega}_k,\widehat{\vu}_k),\quad
\widehat{\vomega}_k\equiv\nabla\times \widehat{\vu}_k;\quad
\widehat{\vomega}_k, \widehat{\vu}_k\in \mathbb{H}\cap\mathbb{O}(1),\ k=0,1,2,\dots
\label{basic-4aa-0}
\end{eqnarray}
where all $\widehat{\vu}_k$ represent  unknown functions. Here we formulate the result. The truncated general
solution $\widehat{\vomega}^{[3]}$ is
\begin{eqnarray}
&&\widehat{\vomega}^{[3]}=\widehat{\vomega}_0+\varepsilon\widehat{\vomega}_1+\varepsilon^2\widehat{\vomega}_2+
\varepsilon^3\widehat{\vomega}_3
\label{gs-0-v}
\end{eqnarray}
The bar-parts $\overline{\vomega}_k$ satisfy the equations
\begin{eqnarray}
&&\overline{\vomega}_{0s}+[\,\overline{\vomega}_0,\overline{\vu}_0]=0\label{eqn-1-0-bar-h-f-v}\\
&&\overline{\vomega}_{1s}+ [\overline{\vomega}_0,\overline{\vu}_1+\overline{\vV}_0]
+[\overline{\vomega}_1,\overline{\vu}_0]=0
\label{eqn-3-bar1-h-f-v}\\
&&\overline{\vomega}_{2s}+[\overline{\vomega}_2,\overline{\vu}_0]+[\overline{\vomega}_1,(\overline{\vu}_1+
\overline{\vV}_0)]
+ [\overline{\vomega}_0,(\overline{\vu}_2+\overline{\vV}_{10}+\overline{\vV}_1+\overline{\vV}_{\chi})] =
\label{eqn-3-bar1-h-f-v}\\
&&\frac{\partial}{\partial x_i}\left(
\overline{\chi}_{ik}\frac{\partial\overline{\omega}_{0i}}{\partial x_k}\right)-
\overline{b}_{ikl}\frac{\partial\overline{\omega}_{0l}}{\partial x_k}-\overline{a}_{ik}\overline{\omega}_{k}
\nonumber
\end{eqnarray}
All these equations are complemented by the constraints $\overline{\vomega}_k\equiv\nabla\times
\overline{\vu}_k$ for $k=0,1,2$.
After (\ref{eqn-1-0-bar-h-f-v})-(\ref{eqn-3-bar1-h-f-v}) are solved, the tilde-parts $\widetilde{\vomega}_k$
of (\ref{gs-0}) are given by the recurrent expressions
\begin{eqnarray}
&&\widetilde{\vomega}_0\equiv 0,\label{sol-0-0-h-f-v}\\
&& \widetilde{\vomega}_1=[\widetilde{\vu}_0^\tau,\overline{\vomega}_0]\label{sol-1-0-h-f-v}\\
&&\widetilde{\vomega}_{2}=-\widetilde{\vomega}_{1s}^\tau-[\overline{\vomega}_0,\widetilde{\vu}_1^\tau]-
[\widetilde{\vomega}_1^\tau,\overline{\vu}_0]-[\overline{\vomega}_1,\widetilde{\vu}_0^\tau]
-\{[\widetilde{\vomega}_1,\widetilde{\vu}_0]\}^\tau
\label{eqn-2-0-tilde-h-integr-f-v}
\\
&&\widetilde{\vomega}_{3}=-\widetilde{\vomega}_{2s}^\tau-[\overline{\vomega}_0,\widetilde{\vu}_2^\tau]-
[\widetilde{\vomega}_2^\tau,\overline{\vu}_0]-[\overline{\vomega}_2,\widetilde{\vu}_0^\tau]-\label{eqn-3-tilde-h-tau-f-v}\\
&&[\widetilde{\vomega}_1^\tau,\overline{\vu}_1]-[\overline{\vomega}_1,\widetilde{\vu}_1^\tau]-
\{[\widetilde{\vomega}_2,\widetilde{\vu}_0]\}^\tau-
\{[\widetilde{\vomega}_1,\widetilde{\vu}_1]\}^\tau\nonumber
\end{eqnarray}
All these equations are complemented by the constraints $\widetilde{\vomega}_k\equiv\nabla\times
\widetilde{\vu}_k$  for $k=0,1,2$.
Three averaged equations (\ref{eqn-1-0-bar-h-f-v})-(\ref{eqn-3-bar1-h-f-v}) can be written as a single
equation that combines advection, pseudo-diffusion, and other mean-field-type terms (with the error
${O}(\varepsilon^3)$)
\begin{eqnarray}
&&\overline{\vomega}_s+[\overline{\vomega},\overline{\vv}] =
\frac{\partial}{\partial x_k}\left(
\overline{\chi}_{kl}\frac{\partial\overline{\omega}_{0i}}{\partial x_l}\right)-
\overline{b}_{ikl}\frac{\partial\overline{\omega}_{0l}}{\partial x_k}-\overline{a}_{ik}\overline{\omega}_{k}
\label{gh-15-v}\\
&&\overline{\vv}\equiv\overline{\vu}_0+\varepsilon(\overline{\vu}_1+\overline{\vV}_0)+\varepsilon^2
(\overline{\vu}_2+\overline{\vV}_{10}+\overline{\vV}_1+\overline{\vV}_{\chi})\label{4.22-v}\\
&&\overline{\kappa}_{ik}=\overline{\chi}_{ik}^{[2]}=\varepsilon^2\overline{\chi}_{ik}\label{4.23-v}\\
&&\overline{\vomega}=\overline{\vomega}^{[2]}=\overline{\vomega}_0+\varepsilon\overline{\vomega}_1+
\varepsilon^2\overline{\vomega}_2
\label{gh-16-v}
\end{eqnarray}
Eqn. (\ref{gh-15-v}) shows that the evolution of $\overline{\vomega}$ represents (with the error
${O}(\varepsilon^3)$) an advection and stretching by  velocity $\overline{\vv}$, pseudo-diffusion with matrix
coefficient $\overline{\kappa}_{ik}=\varepsilon^2\overline{\chi}_{ik}$, and other mean-field deformations. It
is \emph{very} important that $\overline{\vomega}\neq\nabla\times\overline{\vv}$.

One can now write the truncated solution
\begin{eqnarray}
&&\widehat{\vomega}^{[3]}=\overline{\vomega}_0+\overline{\vomega}_1+\widetilde{\vomega}_1+\overline{\vomega}_2+\widetilde{\vomega}_2+
\overline{\vomega}_3+\widetilde{\vomega}_3
\label{trunc-h2-v}
\end{eqnarray}
where all the terms except $\overline{\vomega}_3$ are determined by the above equations. Its mathematical
justification is similar to the one presented in \emph{Sect.}3.

The difference between the results for vortex dynamics and the results for a magnetic field is essential: in
the former case  velocity is prescribed, while (\ref{eqn-1-0-bar-h-f-v})-(\ref{eqn-3-bar1-h-f-v}) represent
the system of equations for unknown velocity and vorticity. The only exception is the main term
$\widetilde{\vu}_0$ of oscillatory velocity, which is potential (due to (\ref{sol-0-0-h-f-v})), and remains
to be prescribed. There are several ways to prescribe it in different problems: (i) it can be forced by
oscillating boundary conditions; (ii) it can appear as self-oscillations; (iii) it can be maintained by an
external oscillating force; or (iv) it can appear in full viscous theory from the procedure of matching an
outer flow with an oscillatory boundary layer solution. Hence two terms of drift velocity $\overline{\vV}_0$
and $\overline{\vV}_1$ (\ref{gs-10}) represent the functions that are `external' for the averaged equations
(\ref{eqn-1-0-bar-h-f-v})-(\ref{eqn-3-bar1-h-f-v}), while all other terms of drift velocity and all
mean-fields (including pseudo-diffusion) are explicitly defined by both $\widetilde{\vu}_0$ and unknown
solutions of (\ref{eqn-1-0-bar-h-f-v}).


\section{Distinguished limit}

We formulate the idea of a distinguish limit using the problem of \emph{Sect}.3 (scalar transport) for a
special ($s$-independent) non-degenerated velocity field
\begin{eqnarray}
&&\vu=\widehat{\vu}(\vx,\tau),\quad\widehat{\vu}\in\mathbb{O}(1)\cap\mathbb{H},\quad\tau\equiv\sigma
t,\quad\varepsilon\equiv 1/\sigma\label{scalar-0-4-vel-DL}
\end{eqnarray}
Let us select the class of \emph{trial} slow time-variables  as $s=\sigma^{\alpha} t$, where $\alpha=\const$
(it should be $\alpha<1$  since $s$-variable must be `slow' in comparison with $\tau=\sigma t$). Then the
two-timing form of eqn. (\ref{scalar-0-4}) is
\begin{eqnarray}\label{scalar-TT-new-DL}
&&\left(\frac{\partial}{\partial\tau}+\varepsilon^{1-\alpha}\frac{\partial}{\partial
s}\right)\widehat{{c}}+\varepsilon\widehat{\vu}\cdot\nabla\widehat{{c}}=0,\quad
\widehat{c}=\widehat{c}(\vx,s,\tau)
\end{eqnarray}
where $\tau$ and $s$ are independent variables. Surprisingly, the slow time-scale (and the value of $\alpha$)
is uniquely determined by the structure of the equation. It is determined as the distinguished limit, which
includes the following properties: (i) the solution for $\alpha=\alpha_d$ is given by a valid asymptotic
procedure; (ii) all solutions for $\alpha_d<\alpha<1$ contain terms secular in $s$, and (iii) for any
$\alpha<\alpha_d$ the system of equations for successive approximations contains internal contradictions and
it is unsolvable (unless it degenerates). One can show that in all three problems, considered in
\emph{Sects.}3-5, $\alpha_d=0$. We have fully studied the first three approximations for all three problems,
therefore it can be accepted that the validity of the procedures with  $\alpha_d=0$ (item (i) above) has been
established. The properties (ii) and (iii) for $\alpha_d=0$ we accept without proof. The value $\alpha_d=0$
justifies our choice $T=T_i$ in (\ref{scalar-0-2}).

We should emphasise that the nature of time-variables $t$ and $s$ is different: the `modulation' variable $t$
in $\widehat{\vu}(\vx,t,\omega t)$ (\ref{scalar-0-5}) is given by the original physical formulation of the
problem, while slow time-variable $s$ in the equation (\ref{scalar-main-eq}) uniquely appears from the
distinguished limit consideration. All considerations of
\emph{Sects.}3-5 are based on the coincidence (which is the same as $\alpha_d=0$)
of slow time-variable $s$ with  physical time $t$. This lucky coincidence allows us to return from a special
formula for velocity (\ref{scalar-0-4-vel-DL}) to the original general functional form (\ref{scalar-0-5}). In
more general situations (when the scales $s$ and $t$ have different orders) one should consider the problems
with three time-variables: $s$, $t$, and $\tau$. In our case $s=t$,  therefore we do not need to use the
three-timing method.

One more example of a distinguished limit can be seen in the averaged equations (\ref{gs-7})-(\ref{gs-9}).
Let us  take a degenerated (predominantly oscillating) flow $\overline{\vu}_0\equiv 0$,
$\widetilde{\vu}_0\not\equiv 0$  and look for a solution with $\alpha=0$. In this case
eqns.(\ref{gs-7}),(\ref{gs-8}) give
\begin{eqnarray}
&&\overline{{c}}_{0s}=0\label{gs-7d}\\
&&\overline{{c}}_{1s}=-(\overline{\vu}_1+\overline{\vV}_0)\cdot\nabla\overline{{c}}_0\label{gs-8d}
\end{eqnarray}
The integration of the first equation gives $\overline{c}_{0}$ as an arbitrary function of $\vx$ only; its
substitution into the second equation shows that (in the generic case) $\overline{c}_{1}$  grows linearly
with $s$. The roots of this trouble lie in the fact that for  predominantly oscillating flows $\alpha_d=-1$
and $s=t/\sigma$, see \cite{VladimirovX2}; a wrongly chosen variable $s$ inevitably leads to the appearance
of secular terms.

\section{Conditional isomorphism between the high-frequency asymptotic solutions and small-amplitude asymptotic solutions}

The area of applicability of the derived above results is essentially expanded by the isomorphism between the
high-frequency asymptotic solutions and the small-amplitude asymptotic solutions. In order to clarify the
existence of such isomorphism, let us rewrite eqn.(\ref{scalar-0-4}) for a newly defined unmodulated velocity
$\vu$
\begin{eqnarray}
&&\vu=\varepsilon_1\widehat{\vu}_1(\vx,\tau),\quad\widehat{\vu}_1\in\mathbb{O}(1)\cap\mathbb{H},\quad\tau\equiv\sigma
t,\quad\sigma=O(1)\label{scalar-0-4-vel}\\
&&{c}_{\tau}+\varepsilon_1\frac{1}{\sigma}\widehat{\vu}_1\cdot\nabla{c}=0,\quad
\Div\widehat{\vu}_1=0,\quad\varepsilon_1\to 0\label{scalar-0-4-D}
\end{eqnarray}
where $\varepsilon_1$ is a small amplitude of  velocity $\vu$. The most important difference with the case of
\emph{Sects.}3-5 is  $\sigma=O(1)$. The two-timing form of solution is
\begin{eqnarray}\label{scales-1}
&&c=\widehat{c}(\vx,s_1,\tau_1)\quad \text{where}\quad\tau_1=\tau;  \quad  s_1=\varepsilon_1^\beta\tau;\quad
\frac{\partial}{\partial\tau}=\frac{\partial}{\partial\tau_1}+\varepsilon_1^\beta\frac{\partial}{\partial s_1}
\end{eqnarray}
where $s_1$ is a \emph{trial} slow time-variable with an indefinite constant $\beta>0$. The substitution of
(\ref{scales-1}) into (\ref{scalar-0-4-D}) gives
\begin{eqnarray}\label{scalar-TT-new}
&&\left(\frac{\partial}{\partial\tau_1}+\varepsilon_1^\beta\frac{\partial}{\partial
s_1}\right)\widehat{{c}}+\varepsilon_1\frac{1}{\sigma}\widehat{\vu}_1\cdot\nabla\widehat{{c}}=0
\end{eqnarray}
which is our stage to declare that the variable $\tau_1$ and $s_1$ are temporarily independent (\emph{cf.}
with (\ref{scalar-0-9})). One can see that the mathematical form of eqn.(\ref{scalar-TT-new}) coincides with
(\ref{scalar-TT-new-DL}). Therefore, according to the result of the previous section, the distinguished limit
corresponds to $\beta=1$ and the final equation to be solved is
\begin{eqnarray}\label{scalar-TT-new-F}
&&\widehat{c}_{\tau_1}+\varepsilon_1
\widehat{c}_{s_1}+\varepsilon_1\frac{1}{\sigma}\widehat{\vu}_1\cdot\nabla\widehat{{c}}=0
\end{eqnarray}
The comparison between (\ref{scalar-TT-new-F}) and  (\ref{scalar-main-eq}) shows the mathematical isomorphism
between the high-frequency asymptotic problem and the small-amplitude asymptotic problem, where the
replacements
\begin{eqnarray}\label{isomorf}
\tau\leftrightarrow\tau_1,\quad s\leftrightarrow s_1,\quad \widehat{\vu}\leftrightarrow\widehat{\vu}_1/\sigma,
\quad\varepsilon\leftrightarrow\varepsilon_1
\end{eqnarray}
provide the transformation of the equations (\ref{scalar-TT-new-F}) and  (\ref{scalar-main-eq}) into each
other.

We need to add some further comments:

1. The slow variable $s_1=\varepsilon_1\sigma t$ is different from $t$, therefore, in contrast with
(\ref{scalar-0-5}), $t$ can not included  to the list of original variables of velocity
(\ref{scalar-0-4-vel}).

2. The area of applicability of the introduced isomorphism is restricted by velocity fields in the form
(\ref{scalar-0-4-vel}). Physically, the expressions in the form (\ref{scalar-0-4-vel}) mean that we are
allowed to consider only  steady velocity fields $\overline{\vu}(\vx)$ with superimposed unmodulated
oscillations $\widetilde{\vu}(\vx,\tau)$. Such functional form can be deliberately chosen for the first two
problems (scalar transport and kinematic MHD). For vortex dynamics one should impose the same functional
restrictions on $\widehat{\vu}_0$.

3. The existence of similar to (\ref{isomorf}) isomorphisms should be checked separately  for each new
equation. In particular, the adding of diffusivity or viscosity can alter the appearance of isomorphism.

4. One can find that the discussed isomorphism can be also established by the simultaneous rescaling of
time-variable $\tau$ and velocity $\vu$ in (\ref{scalar-0-4-vel}). We think that our way is preferable since:
(i) it preserves the original physical variables; (ii) it can be used in more general equations.

5. The high-frequency problems and  small-amplitude problems are physically different. The introduced
isomorphism shows only their mathematical equivalence, which is valid under the formulated restrictions.

6. The existence of isomorphism  (\ref{isomorf}) explains why some high-frequency solutions and
small-amplitude solutions can coincide (or can be close to each other). For example, it explains why the
average equation for vorticity by
\cite{CraikLeib} is very similar to eqn.(\ref{eqn-3-bar1-h-f-v}) and represents a steady version of the
MHD-Drift equation from \cite{VladimirovX2}, taken without magnetic field.

\section{On the nature of pseudo-diffusion (PD)}

The pseudo-diffusion (PD) term that appears in the averaged equations (\ref{gs-9}), (\ref{gs-15}) might be
seen  as a physically new result. The prefix \emph{pseudo-} appears in our study due to three reasons: (i) in
the successive approximations PD appears for the first time in the equation of the second approximation (not
in the zero-order equation); hence, mathematically PD always plays a part of the known RHS in the equation of
the second approximation; (ii) we use only regular asymptotic expansions (\ref{scalar-2}); and (iii) PD
coefficient (matrix) has a specific form of Lie-derivative (\ref{gs-9}),(\ref{gs-14}). The first two reasons
are very common (see \emph{e.g.} \cite{Pedley}), they comply with the role of PD as an effective physical
diffusion. However, the third reason  might lead us to the suggestion that this term gives a $s$-dependent
kinematic deformation of the averaged field, which is not related to physical diffusion. We illustrate the
role of this term in a simple example of three-dimensional translational rigid-body oscillations.

Let an infinite fluid oscillate translationally as a rigid-body with a small displacement
$\widetilde{\vx}(s,\tau)=O(\varepsilon)$, where $\varepsilon=1/\sigma$. Then the Eulerian coordinate of a
particle is $\vx=\overline{\vx}+\widetilde{\vx}$. The drift velocity identically vanishes in all orders
$\overline{\vV}\equiv 0$, hence for any particle $\overline{\vx}\equiv\const$. Therefore, we accept that
$\widetilde{\vx}(0,0)=0$, which means that the Lagrangian (initial) coordinate of each particle coincides
with its unperturbed position $\vX=\overline{\vx}$. Then the solution is
\begin{equation}\label{mm1}
\widehat{c}(\vx,s,\tau)=c^L(\vx-\widetilde{\vx})
\end{equation}
where $c^L(\vX)$ is a given time-independent distribution in Lagrangian coordinates. One can expand both
sides of (\ref{mm1}) as
\begin{equation}\label{mm2}
\widehat{c}_0+\varepsilon\widehat{c}_1+\varepsilon^2\widehat{c}_2+\ldots=
c^L(\vx)-\widetilde{x}_{i}\frac{\partial c^L(\vx)}{\partial x_i} +
\frac{\varepsilon^2}{2}\widetilde{x}_{i}\widetilde{x}_{k}\frac{\partial^2 c^L(\vx)}{\partial x_i\partial x_k}
+\ldots
\end{equation}
where the LHS corresponds to the decomposition (\ref{scalar-2}) with
$\widehat{c}_n=\widehat{c}_n(\vx,s,\tau)=
\overline{c}_n(\vx,s)+\widetilde{c}_n(\vx,s,\tau)$, while the RHS represents the
Taylor's series. The averaging (\ref{oper-1}) of (\ref{mm2}) yields
\begin{equation}\label{mm3}
\overline{c}_0+\varepsilon\overline{c}_1+\varepsilon^2\overline{c}_2+\ldots=
c^L(\vx)+
\frac{\varepsilon^2}{2}\langle\widetilde{\xi}_{i}
\widetilde{\xi}_{k}\rangle\frac{\partial^2 c^L(\vx)}{\partial x_i\partial x_k} +\ldots
\end{equation}
where we have changed $\widetilde{\vx}(s,\tau)$ to $\varepsilon\widetilde{\vxi}(s,\tau)$ (\ref{gs-10}), which
is valid with the given precision (by an elementary consideration one can derive that
$\widetilde{\vx}=\varepsilon\widetilde{\vxi}-\varepsilon^2\widetilde{\vxi}^\tau_s+O(\varepsilon^3)$). The
$s$-derivative of (\ref{mm3}) is
\begin{equation}\label{mm4}
\overline{c}_{0s}+\varepsilon\overline{c}_{1s}+\varepsilon^2\overline{c}_{2s}+\ldots=
\frac{\varepsilon^2}{2}
\langle\widetilde{\xi}_{i}\widetilde{\xi}_{k}\rangle_s\frac{\partial^2 \overline{c}_0(\vx)}{\partial x_i\partial x_k}
+\ldots
\end{equation}
where we have used $c^L=\overline{c}_{0}$ (which is the leading term of (\ref{mm3})). One can see that
eqn.(\ref{mm4}) coincides with (\ref{gs-7})-(\ref{gs-9}) taken for $\overline{\vu}_0\equiv 0$ and for the
zero drift.

This example indicates that PD represents a purely kinematic correction which appears due to the averaging
operation. In addition, one might put forward a conjecture that all mean-field terms in
(\ref{gs-15}),(\ref{gh-15}),(\ref{gh-15-v}) are also purely kinematic and do not represent any physical
processes. However, the same `philosophical' question arises in the theory of turbulence: let us take an
ensemble of flows, which differ from each other only by translations as a whole; then the averaging operation
with a time-dependent distribution function leads to `turbulent diffusion'. Can it be considered physically
meaningful? The answer to this question requires further analysis. Anyway, the significance of PD (and other
mean-field terms) for the results of this paper lies in the discovery of the universal mathematical structure
of the terms, that appear from the straightforward calculations in addition to the drift-type corrections of
the averaged equations.

\section{Discussion}

1. The main novel element of our consideration is a universal structure of averaged equations describing
different processes in non-degenerated oscillatory flows. The most useful  problem (for the analysis of this
structure) is the one for a scalar admixture (\emph{Sect.}3), where the calculations are the simplest and at
the same time sufficient enough to produce the analytical expressions and constructions, that appear in all
other cases, including predominantly oscillating (degenerated) flows, see
\cite{VladimirovX1,VladimirovX2}.

2. The averaged equations for a `frozen-in' magnetic field and vorticity dynamics formally coincide; the
difference between two cases is the constraint linking velocity with vorticity in the latter one. However,
the presence of such a constraint has entirely changed the nature of the equations: the velocity field is
prescribed in the MHD-kinematics but it represents an unknown function in vortex dynamics.

3. For all considered non-degenerated flows a non-linear averaged equation appears only in the mean (zero)
approximation of vortex dynamics; its form coincides with that of the original (exact) equations. All other
equations are linear. The most impressive in the analytical beauty and demanding physical explanations is an
appearance of Lie-derivatives in the pseudo-diffusion matrix and in other mean-fields.

4. In this paper we intensely use the notion of a drift. It is known that a drift velocity can appear from
Lagrangian, Eulerian, or hybrid (Euler-Lagrange) considerations. In our study we use the
\emph{Eulerian drift}, which appears as a result of the Eulerian averaging of the governing PDEs without direct
addressing the motion of particles, see
\cite{CraikLeib, Craik0, Riley, VladimirovX1, VladimirovX2, Ilin}. More classical \emph{Lagrangian drift} appears as the
average motion of Lagrangian particles and its theory is based on the averaging of ODEs, see \cite{Stokes,
Lamb, LH, Batchelor}; the hybrid drift coincides with the Lagrangian one, see
\cite{ McIntyre, Craik0, Soward2, VladimirovX1}. It is known that Lagrangian and Eulerian drifts are
similar, but not identical to each other, see
\cite{VladimirovX1}.

5. The first two examples (the transport of a scalar and the kinematic MHD) are not always mutually
independent: for the translationally-invariant motions eqn.(\ref{exact-1-h}) can be reduced to
(\ref{scalar-0-1}), see \cite{VladMof}.

6. The small parameter of our asymptotic theory is  inverse dimensionless frequency $\varepsilon=1/\omega$.
One can see that (since the oscillatory velocity is $O(1)$) the dimensionless characteristic oscillatory
displacement of a material particle is $O(\varepsilon)$. It is interesting, that for the isomorphic
small-amplitude flow of \emph{Sect.}7 this displacement is $O(\varepsilon_1)$.

7. For a finite and time-dependent flow domain $\mathcal{D}(t)$ the definition of average (\ref{oper-1})
works only if $\vx\in\mathcal{D}$ at any instant. If it is not true, then the theory should include the
`projection' of a boundary condition on an `undisturbed' boundary. Such a consideration requires the
smallness of the oscillatory displacements of fluid particles, which are always proportional to $\varepsilon$
(see item 6 above). Hence, the generalizations of all our results to finite domains can be undertaken without
adding any new small parameters to the list (\ref{scalar-0-8}).

8. Viscosity or diffusivity can be routinely added to all the above considerations as the RHS-term
$\kappa\nabla^2\widehat{a}$, $\nu\nabla^2\widehat{\vu}$, \emph{etc.} However, the final appearance of such
terms depends on the order of magnitude of the new small parameter (like $1/Pe\equiv\kappa/\sigma L^2$, where
$Pe$ is the Peclet number) in terms of $\varepsilon$. Different relations between small parameters will
produce different averaged equations.  The most `harmless' case of adding  viscosity or diffusivity is to
take $\kappa/\sigma L^2=O(\varepsilon^2)$. In this case  viscous or diffusion terms will be just added to all
the equations of the second approximation. This assumption has been intensely used in  biological
applications, see
\cite{Pedley}. It is interesting to note that in such cases  dimensionless physical diffusion and
pseudo-diffusion are of the same order of magnitude. We avoid viscosity and diffusivity in this paper in
order to focus our attention on `pure' kinematics and dynamics.

9. The incorporation of  density stratification and/or compressibility into all considered problems is
straightforward. In fact, a similar theory of the predominantly oscillating flows was developed for a
compressible fluid, see \cite{VladimirovX1}.

10. In this paper we consider periodic (in  fast time-variable $\tau$) functions. The studies of non-periodic
in $\tau$ oscillations can represent the next stages of research. For example, such a development has been
already available for Langmuir circulations, see \cite{CraikLeib, Leibovich, Craik0}.

11. The described transformations make partially visible the same properties of the governing equations of
fluid dynamics as were studied by the hybrid Euler-Lagrange GLM-procedure, see \cite{McIntyre, Buhler,
Soward2}. However, the GLM-method in its original form is not adapted to the two-timing framework: it is
sufficient to note that a fast-time variable and a related small parameter (the ratio of two time-scales) do
not appear in
\cite{McIntyre}. The two-timing adaptation of the GLM-kinematics has been developed by \cite{VladimirovX1}.
Further steps in this direction are required to introduce the two-timing dynamics.

12. Degenerated flows with $\overline{\vu}_0\equiv 0$ (predominantly oscillating flows) were considered by
\cite{Stokes,CraikLeib,Riley,Pedley,Buhler,VladimirovX1,Her,VladimirovX2}, and  many others.
The results of \cite{VladimirovX1,VladimirovX2} show that the averaged governing equations of degenerated
flows contain the same (or similar) building blocks/elements, as the ones presented above. The vortex
dynamics of such flows is extremely interesting theoretically, since it produces non-linear equations for
averaged vorticity, that include a drift in their coefficients. A classical application of this theory is the
describing of the Langmuir circulations, see \cite{CraikLeib,Leibovich,Thorpe}.

13. There are many areas where the results of \emph{Sects.}3-7 could be applied, such as waves in a flow with
strong shear, flows past translationally moving and oscillating solids, oscillatory flows in channels and
pipes, various acoustical problems, \emph{etc.} The number of applications can be very large. However we do
not consider any applications, since: (i) the focus of this paper is the universal structure of the averaged
governing equations; (ii) every application of a general theory requires its own paper.

14.  This paper is written as Part 1 of a series, where  Part 2 (a similar theory for degenerated flows) is
based on \cite{VladimirovX1}. A number of examples illustrating the values of drift and pseudo-diffusion is
given in Part 2, since predominantly oscillating flows have many well-developed applications.

\begin{acknowledgments}
The author is grateful to Profs. A.D.D.Craik, H.K.Moffatt, T.J.Pedley, and A.M.Soward for helpful
discussions. The author thanks the Department of Mathematics of the University of York for the
research-stimulating environment.
\end{acknowledgments}

\appendix

\section{Deriving of the equations for a passive admixture \label{sect04a}}
The equation of zero approximation (\ref{scalar-r-1}) is
\begin{eqnarray}
&&\widehat{c}_{0\tau}=\widetilde{c}_{0\tau}=0\label{scalar-3}
\end{eqnarray}
Its unique solution is $\widetilde{c}_{0}\equiv 0$ (\ref{f-tilde=0}), while $\overline{c}_0$ remains
undetermined; let us write it as
\begin{eqnarray}
&&\widetilde{{c}}_0\equiv 0,\quad \overline{{c}}_0=\boxed{?}\label{scalar-4}
\end{eqnarray}
where the first equality gives (\ref{gs-1}). The equation of the first approximation (\ref{scalar-r-2}) (with
the use of (\ref{scalar-4})) is:
\begin{eqnarray}
&&\widetilde{{c}}_{1\tau}+\widehat{\vu}_0\cdot\nabla \overline{{c}}_0+\overline{{c}}_{0s}=0\label{scalar-5}
\end{eqnarray}
Its bar-part (\ref{oper-3}) and tilde-part (\ref{oper-2}) are
\begin{eqnarray}
&&\overline{{c}}_{0s}+\overline{\vu}_0\cdot\nabla \overline{{c}}_0=0\label{scalar-6}\\
&&\widetilde{{c}}_{1\tau}+\widetilde{\vu}_0\cdot\nabla\overline{{c}}_0=0\label{scalar-7}
\end{eqnarray}
where (\ref{scalar-6}) represents the final equation for $\overline{{c}}_0$ (\ref{gs-7}). The
$\mathbb{T}$-integration (\ref{oper-7}) of (\ref{scalar-7}) yields (\ref{gs-2}) and keeps $\overline{{c}}_1$
unknown
\begin{eqnarray}
&&\widetilde{{c}}_1=-\widetilde{\vu}_0^\tau\cdot\nabla\overline{{c}}_0,\quad
\overline{{c}}_1=\boxed{?}\label{scalar-8}
\end{eqnarray}
The equation of second approximation (\ref{scalar-r-3}) that takes into account (\ref{scalar-4}) is
\begin{eqnarray}
&&\widetilde{{c}}_{2\tau}+\widehat{\vu}_1\cdot\nabla \overline{{c}}_0+
\widehat{\vu}_0\cdot\nabla\widehat{{c}}_1+\widehat{{c}}_{1s}=0\label{scalar-9}
\end{eqnarray}
Its bar-part and tilde-part are
\begin{eqnarray}
&&\overline{{c}}_{1s}+\overline{\vu}_1\cdot\nabla\overline{{c}}_0+
\overline{\vu}_0\cdot\nabla\overline{{c}}_1
+\langle\widetilde{\vu}_0\cdot\nabla\widetilde{c}_1\rangle=0\label{scalar-10}\\
&&\widetilde{{c}}_{2\tau}+\widetilde{\vu}_1\cdot\nabla\overline{{c}}_0+
\overline{\vu}_0\cdot\nabla\widetilde{{c}}_1+\widetilde{\vu}_0\cdot\nabla\overline{{c}}_1+
\{\widetilde{\vu}_0\cdot\nabla\widetilde{{c}}_1\}+\widetilde{{c}}_{1s}=0
\label{scalar-11}
\end{eqnarray}
where $\langle\cdot\rangle$  stands for the average (\ref{oper-1}) and $\{\cdot\}$ for the tilde-part
(\ref{oper-5}). Eqn. (\ref{scalar-10}) can be transformed (with the use of (\ref{scalar-8}) and
(\ref{commutr})-(\ref{oper-15})) into the form (\ref{gs-8})
\begin{eqnarray}
&&\overline{{c}}_{1s}+\overline{\vu}_0\cdot\nabla\overline{{c}}_1+
(\overline{\vu}_1+\overline{\vV}_0)\cdot\nabla\overline{{c}}_0 =0
\label{scalar-12}\\
&&\overline{\vV}_0\equiv\frac{1}{2}\langle[\widetilde{\vu}_0,\widetilde{\vu}_0^\tau]\rangle\label{drift-vel0-0-h}
\end{eqnarray}
while the $\mathbb{T}$-integration (\ref{oper-7}) of (\ref{scalar-11}) gives (\ref{gs-3})
\begin{eqnarray}
&&\widetilde{{c}}_{2}=-\widetilde{\vu}_1^\tau\cdot\nabla\overline{{c}}_0-
\overline{\vu}_0\cdot\nabla\widetilde{{c}}_1^\tau-\widetilde{\vu}_0^\tau\cdot\nabla\overline{{c}}_1
-\{\widetilde{\vu}_0\cdot\nabla\widetilde{{c}}_1\}^\tau-\widetilde{{c}}_{1s}^\tau,\quad
\overline{{c}}_2=\boxed{?}
\label{scalar-13}
\end{eqnarray}
The equation of third approximation (\ref{scalar-r-3a}) that takes into account (\ref{scalar-4}) is
\begin{eqnarray}
&&\widetilde{{c}}_{3\tau}+\widehat{\vu}_2\cdot\nabla \overline{{c}}_0+
\widehat{\vu}_0\cdot\nabla\widehat{{c}}_2+\widehat{\vu}_1\cdot\nabla\widehat{{c}}_1+
\widehat{{c}}_{2s}=0\label{scalar-14}
\end{eqnarray}
Its bar-part and tilde-part are
\begin{eqnarray}
&&\overline{{c}}_{2s}+\overline{\vu}_2\cdot\nabla\overline{{c}}_0+
\overline{\vu}_0\cdot\nabla\overline{{c}}_2+\overline{\vu}_1\cdot\nabla\overline{{c}}_1
+\langle\widetilde{\vu}_0\cdot\nabla\widetilde{c}_2\rangle
+\langle\widetilde{\vu}_1\cdot\nabla\widetilde{c}_1\rangle
=0\label{scalar-15}\\
&&\widetilde{{c}}_{3\tau} +\widetilde{\vu}_2\cdot\nabla\overline{{c}}_0+
\widetilde{\vu}_0\cdot\nabla\overline{{c}}_2+
\overline{\vu}_0\cdot\nabla\widetilde{{c}}_2+
\widetilde{\vu}_1\cdot\nabla\overline{{c}}_1+
\overline{\vu}_1\cdot\nabla\widetilde{{c}}_1+\label{scalar-16}\\
&&\{\widetilde{\vu}_0\cdot\nabla\widetilde{{c}}_2\}+
\{\widetilde{\vu}_1\cdot\nabla\widetilde{{c}}_1\}
+\widetilde{{c}}_{2s}=0\nonumber
\end{eqnarray}
We transform the `Reynolds stress' terms in (\ref{scalar-15}) with the use of
(\ref{scalar-8}),(\ref{scalar-13}),(\ref{commutr})-(\ref{oper-15})
\begin{eqnarray}
&&R_c\equiv\langle\widetilde{\vu}_1\cdot\nabla\widetilde{c}_1\rangle+
\langle\widetilde{\vu}_0\cdot\nabla\widetilde{c}_2\rangle=R_{c1}+R_{c2}+R_{c3}+R_{c4}\label{scalar-16a}\\
&&R_{c1}\equiv-\langle(\widetilde{\vu}_0\cdot\nabla)(\widetilde{\vu}^{\tau}_0\cdot\nabla)\rangle\overline{c}_1=
\overline{\vV}_0\cdot\nabla\overline{c}_1\\
&&R_{c2}\equiv-\langle(\widetilde{\vu}_1\cdot\nabla)(\widetilde{\vu}^{\tau}_0\cdot\nabla)\rangle\overline{c}_0
-\langle(\widetilde{\vu}_0\cdot\nabla)(\widetilde{\vu}^{\tau}_1\cdot\nabla)\rangle\overline{c}_0 =
\overline{\vV}_{01}\cdot\nabla\overline{c}_0\\
&&R_{c3}\equiv-\langle(\widetilde{\vu}_0\cdot\nabla)\{(\widetilde{\vu}_0\cdot\nabla)\widetilde{c}_1\}^\tau\rangle
=\langle(\widetilde{\vu}_0^\tau\cdot\nabla)(\widetilde{\vu}_0\cdot\nabla)\widetilde{c}_1\rangle=\\
&&-\langle(\widetilde{\vu}_0^\tau\cdot\nabla)(\widetilde{\vu}_0\cdot\nabla)(\widetilde{\vu}_0^\tau\cdot\nabla)\rangle
\widetilde{c}_0=\overline{\vV}_{1}\cdot\nabla\overline{c}_0\nonumber
\\
&&R_{c4}\equiv-\langle(\widetilde{\vu}_0\cdot\nabla)\widetilde{c}_{1s}^\tau\rangle
-\langle(\widetilde{\vu}_0\cdot\nabla)(\overline{\vu}_0\cdot\nabla)\widetilde{c}_1^\tau\rangle=\label{scalar-16b}\\
&&\langle(\widetilde{\vu}_0^\tau\cdot\nabla)D_0\widetilde{c}_{1}\rangle=
\overline{\vV}_{\chi}\cdot\nabla\overline{{c}}_0 -\frac{\partial}{\partial x_i}\left(
\overline{\chi}_{ik}\frac{\partial\overline{c}_0}{\partial x_k}\right)\nonumber\\
&&\overline{\vV}_0\equiv
\frac{1}{2}\langle[\widetilde{\vu}_0,\widetilde{\vxi}]\rangle,\quad
\overline{\vV}_1\equiv\frac{1}{3}\langle[[\widetilde{\vu}_0,\widetilde{\vxi}],\widetilde{\vxi}]\rangle,\quad
\widetilde{\vxi}\equiv\widetilde{\vu}_0^\tau
\label{scalar-16c}\\
&&\overline{\vV}_{10}\equiv\langle[\widetilde{\vu}_1,\widetilde{\vxi}]\rangle,
\quad
\vV_\chi\equiv\frac{1}{2}\langle[\widetilde{\vxi},\mathfrak{L}\widetilde{\vxi}]\rangle,\quad
\overline{\chi}_{ik}\equiv\frac{1}{2}
\mathfrak{L}\langle\widetilde{\xi}_i\widetilde{\xi}_k\rangle,\label{scalar-16-d}
\end{eqnarray}
where the operator $\mathfrak{L}$ (Lie-derivative) is such that its action on $\widetilde{\vxi}$ and
$\langle\widetilde{\xi}_i\widetilde{\xi}_k\rangle$ is defined as
\begin{eqnarray}
&&\mathfrak{L}\widetilde{\vxi}\equiv\widetilde{\vxi}_s+[\widetilde{\vxi},\overline{\vu}_0]\label{scalar-16-e}\\
&&\mathfrak{L}\langle\widetilde{\xi}_i\widetilde{\xi}_k\rangle\equiv\left(\partial_s+
\overline{\vu}_0\cdot\nabla\right)\langle\widetilde{\xi}_i\widetilde{\xi}_k\rangle
-\frac{\partial\overline{u}_{0k}}{\partial x_m}\langle\widetilde{\xi}_i\widetilde{\xi}_m\rangle-
\frac{\partial\overline{u}_{0i}}{\partial x_m}\langle\widetilde{\xi}_k\widetilde{\xi}_m\rangle
\label{scalar-16-f}
\end{eqnarray}
Hence eqn.(\ref{scalar-15}) takes a form
\begin{eqnarray}
&&\overline{{c}}_{2s}+\overline{\vu}_0\cdot\nabla\overline{{c}}_2+
(\overline{\vu}_1+\overline{\vV}_0)\cdot\nabla\overline{{c}}_1+
\label{scalar-17}\\
&&(\overline{\vu}_2+\overline{\vV}_{10}+\overline{\vV}_1+\overline{\vV}_{\chi})\cdot\nabla\overline{{c}}_0
=\frac{\partial}{\partial x_i}\left(
\overline{\chi}_{ik}\frac{\partial\overline{a}_0}{\partial x_k}\right)
\nonumber
\end{eqnarray}
while (\ref{scalar-16}) can be $\mathbb{T}$-integrated (\ref{oper-7})
\begin{eqnarray}
&&\widetilde{{c}}_{3\tau}=-\widetilde{\vu}_2^{\tau}\cdot\nabla\overline{{c}}_0-
\widetilde{\vu}_0^{\tau}\cdot\nabla\overline{{c}}_2-
\overline{\vu}_0\cdot\nabla\widetilde{{c}}_2^{\tau}-
\widetilde{\vu}_1^{\tau}\cdot\nabla\overline{{c}}_1-
\overline{\vu}_1\cdot\nabla\widetilde{{c}}_1^{\tau}-\label{scalar-18a}\\
&&\{\widetilde{\vu}_0\cdot\nabla\widetilde{{c}}_2\}^{\tau}-
\{\widetilde{\vu}_1\cdot\nabla\widetilde{{c}}_1\}^{\tau}
+\widetilde{{c}}_{2s}^{\tau},\quad
\overline{{c}}_3=\boxed{?}
\nonumber
\end{eqnarray}

\section{Deriving of  equations for  kinematic MHD \label{sect04a}}

 The equations of the zeroth approximation is
\begin{eqnarray}
&&\widehat{\vh}_{0\tau}=\widetilde{\vh}_{0\tau}=0\label{eqn-0-0-h}
\end{eqnarray}
Its unique solution is $\widetilde{\vh}_{0}\equiv 0$ (\ref{f-tilde=0}), while $\overline{\vh}_0$ remains
undetermined; let us write it as
\begin{eqnarray}
&&\widetilde{\vh}_0\equiv 0,\quad \overline{\vh}_0=\boxed{?}\label{sol-0-0-h}
\end{eqnarray}
The equation of the first approximation of (\ref{main-eq-0-h}) (with the use of (\ref{sol-0-0-h})) is:
\begin{eqnarray}
&&\widetilde{\vh}_{1\tau}+[\,\overline{\vh}_0,\widehat{\vu}_0]+\overline{\vh}_{0s}=0\label{eqn-1-0}
\end{eqnarray}
Its bar-part (\ref{oper-3}) and tilde-part (\ref{oper-2}) are
\begin{eqnarray}
&&\overline{\vh}_{0s}+[\,\overline{\vh}_0,\overline{\vu}_0]=0\label{eqn-1-0-bar-h}\\
&&\widetilde{\vh}_{1\tau}+[\,\overline{\vh}_0,\widetilde{\vu}_0]=0\label{eqn-1-0-tilde-h}
\end{eqnarray}
Eqn.(\ref{eqn-1-0-bar-h}) represents a final equation for $\overline{\vh}_0$. The $\mathbb{T}$-integration
(\ref{oper-7}) of (\ref{eqn-1-0-tilde-h}) yields
\begin{eqnarray}
&&\widetilde{\vh}_1=[\widetilde{\vu}_0^\tau,\overline{\vh}_0],\quad
\overline{\vh}_1=\boxed{?}\label{sol-1-0-h}
\end{eqnarray}
The equation of the second approximation that takes into account (\ref{sol-0-0-h}) is
\begin{eqnarray}
&&\widetilde{\vh}_{2\tau}+\widehat{\vh}_{1s}+[\overline{\vh}_0,\widehat{\vu}_1]+
[\widehat{\vh}_1,\widehat{\vu}_0]=0\label{eqn-2-0-h}
\end{eqnarray}
Its bar-part and tilde-part are
\begin{eqnarray}
&&\overline{\vh}_{1s}+[\overline{\vh}_0,\overline{\vu}_1]+ [\overline{\vh}_1,\overline{\vu}_0]
+\langle[\widetilde{\vh}_1,\widetilde{\vu}_0]\rangle=0\label{sol-2-0-bar-h}\\
&&\widetilde{\vh}_{2\tau}+\widetilde{\vh}_{1s}+[\overline{\vh}_0,\widetilde{\vu}_1]+
[\widetilde{\vh}_1,\overline{\vu}_0]+[\overline{\vh}_1,\widetilde{\vu}_0]+\{[\widetilde{\vh}_1,\widetilde{\vu}_0]\}=0
\label{eqn-2-0-tilde-h}
\end{eqnarray}
where $\langle\cdot\rangle$  stands for the average operation (\ref{oper-1}) and $\{\cdot\}$ for the
tilde-part (\ref{oper-5}). The Reynolds-stress-type term  in (\ref{sol-2-0-bar-h}) can be transformed with
the use of (\ref{sol-1-0-h}) and (\ref{commutr})-(\ref{oper-15}) as
\begin{eqnarray}
&&\langle[\widetilde{\vh}_1,\widetilde{\vu}_0]\rangle=
\langle[[\widetilde{\vu}_0^\tau\overline{\vh}_0],\widetilde{\vu}_0]\rangle=
[\overline{\vh}_0,\overline{\vV}_0]\nonumber\\
&&\overline{\vV}_0\equiv\frac{1}{2}\langle[\widetilde{\vu}_0,\widetilde{\vu}_0^\tau]\rangle\label{drift-vel0-0-h}
\end{eqnarray}
Hence (\ref{sol-2-0-bar-h}) takes a form
\begin{eqnarray}
&&\overline{\vh}_{1s}+ [\overline{\vh}_0,\overline{\vu}_1+\overline{\vV}_0]
+[\overline{\vh}_1,\overline{\vu}_0]=0
\label{eqn-3-bar1-h}
\end{eqnarray}
while (\ref{drift-vel0-0-h}) can be $\mathbb{T}$-integrated (\ref{oper-7})
\begin{eqnarray}
&&\widetilde{\vh}_{2}=-\widetilde{\vh}_{1s}^\tau-[\overline{\vh}_0,\widetilde{\vu}_1^\tau]-
[\widetilde{\vh}_1^\tau,\overline{\vu}_0]-[\overline{\vh}_1,\widetilde{\vu}_0^\tau]
-\{[\widetilde{\vh}_1,\widetilde{\vu}_0]\}^\tau,\quad \overline{\vh}_2=\boxed{?}
\label{eqn-2-0-tilde-h-integr}
\end{eqnarray}
The equation of the third approximation that takes into account (\ref{sol-0-0-h}) is
\begin{eqnarray}
&&\widetilde{\vh}_{3\tau}+\widehat{\vh}_{2s}+[\overline{\vh}_0,\widehat{\vu}_2]+
[\widehat{\vh}_2,\widehat{\vu}_0]+[\widehat{\vh}_1,\widehat{\vu}_1]=0\label{eqn-3-h}
\end{eqnarray}
Its bar-part and tilde-part are
\begin{eqnarray}
&&\overline{\vh}_{2s}+[\overline{\vh}_0,\overline{\vu}_2]+ [\overline{\vh}_2,\overline{\vu}_0] +
[\overline{\vh}_1,\overline{\vu}_1]+\langle[\widetilde{\vh}_2,\widetilde{\vu}_0]\rangle
+\langle[\widetilde{\vh}_1,\widetilde{\vu}_1]\rangle=0\label{sol-3-h}\\
&&\widetilde{\vh}_{3\tau}+\widetilde{\vh}_{2s}+[\overline{\vh}_0,\widetilde{\vu}_2]+
[\widetilde{\vh}_2,\overline{\vu}_0]+[\overline{\vh}_2,\widetilde{\vu}_0]+\label{eqn-3-tilde-h}\\
&&[\widetilde{\vh}_1,\overline{\vu}_1]+[\overline{\vh}_1,\widetilde{\vu}_1]+
\{[\widetilde{\vh}_2,\widetilde{\vu}_0]\}+
\{[\widetilde{\vh}_1,\widetilde{\vu}_1]\}=0\nonumber
\end{eqnarray}
We transform the Reynolds-stress-type terms in (\ref{sol-3-h}) with the use of
(\ref{sol-1-0-h}),(\ref{eqn-2-0-tilde-h-integr}),(\ref{commutr})-(\ref{oper-15})
\begin{eqnarray}
&&R_h\equiv\langle[\widetilde{\vh}_1,\widetilde{\vu}_1]\rangle+\langle[\widetilde{\vh}_2,\widetilde{\vu}_0]\rangle=
R_{h1}+R_{h2}+R_{h3}+R_{h4}\label{h-16a}
\\
&&R_{h1}\equiv-\langle[[\overline{\vh}_1,\widetilde{\vu}_0^\tau],\widetilde{\vu}_0]\rangle=
[\overline{\vh}_1,\overline{\vV}_0]\label{h-16aa}
\\
&&R_{h2}\equiv -\langle[[\overline{\vh}_0,\widetilde{\vu}_0^\tau],\widetilde{\vu}_1]\rangle
-\langle[[\overline{\vh}_0,\widetilde{\vu}_1^\tau],\widetilde{\vu}_0]\rangle=[\overline{\vh}_0,\overline{\vV}_{01}]
\label{h-16aab}\\
&&R_{h3}\equiv -\langle[\{[\widetilde{\vh}_1,\widetilde{\vu}_0^\tau]\}^\tau,\widetilde{\vu}_0]\rangle
=[\overline{\vh}_0,\overline{\vV}_{1}]\label{h-16ab}
\\
&&-\langle(\widetilde{\vu}_0^\tau\cdot\nabla)(\widetilde{\vu}_0\cdot\nabla)(\widetilde{\vu}_0^\tau\cdot\nabla)\rangle
\widetilde{c}_0=\overline{\vV}_{1}\cdot\nabla\overline{c}_0
\label{h-16ac}\\
&&R_{h4}\equiv-\langle[\widetilde{\vh}_1^\tau,\widetilde{\vu}_0]\rangle
-\langle[[\widetilde{\vh}_1^\tau,\overline{\vu}_0],\widetilde{\vu}_0]\rangle =
\langle[\mathfrak{L}\widetilde{\vh}_1,\widetilde{\vu}_0^\tau]\rangle=
\label{h-16b}\\
&& [\overline{\vh}_0,\overline{\vV}_{\chi}] -\frac{\partial}{\partial x_k}\left(
\overline{\chi}_{kl}\frac{\partial\overline{h}_{0i}}{\partial x_l}\right)
+2\overline{b}_{ikl}\frac{\partial\overline{h}_{0l}}{\partial x_k}+\overline{a}_{ik}\overline{h}_{k}
\nonumber
\end{eqnarray}
where the additional to (\ref{scalar-16c}),(\ref{scalar-16-d}) notations are:
\begin{eqnarray}
\overline{b}_{ikl}\equiv \mathfrak{L}\left\langle\widetilde{\xi}_k\frac{\partial \widetilde{\xi}_i}
{\partial x_l}\right\rangle,\quad
\overline{a}_{ik}\equiv \mathfrak{L} \left\langle \widetilde{\xi}_l\frac{\partial^2 \widetilde{\xi}_i}
{\partial x_l\partial x_k}  -
\frac{\partial \widetilde{\xi}_l}{\partial x_k}\frac{\partial \widetilde{\xi}_i}{\partial x_l}\right\rangle
\label{h-18}
\end{eqnarray}
Hence eqn.(\ref{scalar-15}) takes a form (with the $i$-component in the LHS)
\begin{eqnarray}
&&\left.\overline{\vh}_{2s}+[\overline{\vh}_2,\overline{\vu}_0]+[\overline{\vh}_1,(\overline{\vu}_1+\overline{\vV}_0)]
+ [\overline{\vh}_0,(\overline{\vu}_2+\overline{\vV}_{10}+\overline{\vV}_1+\overline{\vV}_{\chi})]\right|_i
=\label{eqn-3-bar1-h}
\\
&&\frac{\partial}{\partial x_l}\left(
\overline{\chi}_{lk}\frac{\partial\overline{h}_{0i}}{\partial x_k}\right)-
\overline{b}_{ikl}\frac{\partial\overline{h}_{0l}}{\partial x_k}-\overline{a}_{ik}\overline{h}_{k}
\nonumber
\end{eqnarray}
while (\ref{eqn-3-tilde-h}) can be $\mathbb{T}$-integrated (\ref{oper-7})
\begin{eqnarray}
&&\widetilde{\vh}_{3}=-\widetilde{\vh}_{2s}^\tau-[\overline{\vh}_0,\widetilde{\vu}_2^\tau]-
[\widetilde{\vh}_2^\tau,\overline{\vu}_0]-[\overline{\vh}_2,\widetilde{\vu}_0^\tau]-\label{eqn-3-tilde-h-tau}\\
&&[\widetilde{\vh}_1^\tau,\overline{\vu}_1]-[\overline{\vh}_1,\widetilde{\vu}_1^\tau]-
\{[\widetilde{\vh}_2,\widetilde{\vu}_0]\}^\tau-
\{[\widetilde{\vh}_1,\widetilde{\vu}_1]\}^\tau
,\quad
\overline{{\vh}}_3=\boxed{?}
\nonumber
\end{eqnarray}

\end{document}